\documentclass[aps,10pt,superscriptaddressm,prb,twocolumn]{revtex4}

\usepackage{epsfig}     
\usepackage{dcolumn}    
\usepackage{bm}         

\newcommand{\dd}{\textrm{d}}
\newcommand{\erf}{\mathrm{erf}}
\newcommand{\arccosh}{\mathop{\mathrm{arccosh}}}
\newcommand{\textfoot}[1]{\textrm{\footnotesize{#1}}}
\newcommand{\distvar}{d}

\begin{document}

\title{Multiphase coexistence in polydisperse colloidal mixtures} 
\author{C.~Grodon}
\affiliation{Max-Planck-Institut f{\"u}r Metallforschung,
Heisenbergstr.~3, D-70569 Stuttgart, Germany\\}
\affiliation{Institut f{\"u}r Theoretische und Angewandte Physik,
Universit{\"a}t Stuttgart, Pfaffenwaldring 57, D-70569 Stuttgart, Germany\\}
\author{R.~Roth}
\affiliation{Max-Planck-Institut f{\"u}r Metallforschung,
Heisenbergstr.~3, D-70569 Stuttgart, Germany\\}
\affiliation{Institut f{\"u}r Theoretische und Angewandte Physik,
Universit{\"a}t Stuttgart, Pfaffenwaldring 57, D-70569 Stuttgart, Germany\\}

\begin{abstract}
We study the phase behavior of mixtures of monodisperse colloidal spheres with
a depletion agent which can have arbitrary shape and can possess a
polydisperse size or shape distribution. In the low concentration limit,
considered here, we can employ the free-volume theory and take the geometry of
particles of the depletion agent into account within the framework of
fundamental measure theory. We apply our approach to study the phase diagram
of a mixture of (monodisperse) colloidal spheres and two polydisperse polymer
components. By fine tuning the distribution of the polymer it is possible to
construct a complex phase diagram which exhibits two stable critical points.
\end{abstract}

\maketitle
\section{Introduction}

In the study of the phase behavior of colloidal mixtures there are two very
distinct approaches. From a theoretical point of view one would like to keep
the description of the mixture as simple as possible in order to be able to
calculate the phase diagram using well established theoretical
tools. Furthermore, it is possible to {\em understand} the main physics,
i.e. the phase behavior, bulk correlation functions, or inhomogeneous density
distributions, in simple terms. One example of such an theoretical approach is
the Asakura-Oosawa-Vrij (AOV) model for colloid-polymer 
mixtures.\cite{bib::asakura1954,bib::asakura1958,bib::vrij1976} Within the
AOV model, colloids are treated as hard spheres, while polymer behave like hard
spheres only when interacting with colloids. The polymer-polymer interaction
vanishes. Based on this model, one finds that the phase diagram of a
colloid-polymer mixture, assuming perfectly monodisperse colloids and polymer,
resembles that of a simple fluid, and the reservoir polymer density plays the
role of an inverse temperature. One, relatively simple but successful,
theoretical approach for treating the phase behavior of this model
colloid-polymer mixture is the free-volume theory (FVT) that assumes a low
concentration of the depletion agent, the polymer.\cite{bib::oversteegen2004}

A very different picture of the phase behavior of colloidal mixture originates
from experimental studies. While in some cases it is possible to prepare model
systems with almost monodisperse colloids and 
polymer,\cite{bib::calderon1993,bib::pusey1994} the generic case of a
colloidal sample is polydisperse due to imperfections of the synthesis
process. Polydispersity can be found both in size and in shape
distributions. From the theoretical point of view, polydispersity is often
considered as a defect, because it complicates calculations. However,
polydispersity can give rise to interesting behavior, which can be utilized in
experiments,\cite{bib::fairhurst2004} if the role of polydispersity is
sufficiently well understood. One example is the study of the glass 
transition, where polydispersity in model hard colloidal spheres has the role
as to suppress the freezing transition at high density of colloids so that a
glass can form.\cite{martin03} 

There are various very different ways to achieve a successful description. For
hard and permeable spheres Salacuse and Stell\cite{bib::salacuse1982} showed a
thermodynamic approach based on the Percus-Yevick equation of state and the
more accurate approach by Mansoori {\it et al}. The polydispersity in the
chain-length of self-excluding polymer has been treated in
Ref.~\onlinecite{bib::paricaud2004}. Polydispersity has been incorporated in
the theory in different manner. Near-monodisperse distributions have been
treated in a perturbative fashion\cite{bib::evans2001} and bimodal size
distributions has been studied in Ref.~\onlinecite{bib::warren1999}. 

In the present study we wish to increase the understanding of
effects that can be caused by polydispersity further. To this end we consider a
mixture of monodisperse colloidal spheres with a polydisperse depletion
agent. We keep the theory simple by assuming that the depletion agent is
efficient and can produce a sufficiently strong depletion attraction between
colloidal spheres at rather low concentrations. In this limit we can
generalize the free-volume theory within the framework of fundamental measure
theory\cite{bib::rosenfeld1989} (FMT) to cases of polydisperse size or shape
distributions.

As an application of our general and powerful approach we study the phase
behavior of a ternary mixture of (monodisperse) colloidal spheres with two
polydisperse polymer components with distinct size difference. Our study shows
that the distribution of the depletion agent can be employed to tune the phase
behavior of colloidal mixtures in a wide range and it is possible to generate
a third fluid phase. Beside the low density (in colloids) gas and a high
density liquid, we find a low density liquid phase. This is similar to
findings in systems in which two length scales are introduced in the
inter-particle interaction
potentials.\cite{bib::hemmer1970,bib::stell1972,bib::loverso2003}

The paper is structured as follows. In Section~\ref{sec::theory} we recall
the theoretical basics of free-volume theory within the framework of FMT.
We extend the approach of Ref.~\onlinecite{bib::oversteegen2005} to
treat discrete and continuous distributions of the depletion agent. In
Section~\ref{sec::distributions} we discuss a few distribution functions which
are frequently used in studies of polydisperse systems and show that for
infinitely thin needles or platelets various types of size distributions
lead to explicit expressions for the effective free-volume fraction
$\alpha_{\textrm{eff}}$. In Section~\ref{sec::results} we show and discuss as
an application of our theory the phase diagrams for the case of a ternary
mixture of (monodisperse) colloidal spheres and two polydisperse polymer
components. We discuss the results and conclude in
Section~\ref{sec::discussion}.

\section{Theory}\label{sec::theory}
We study mixtures of colloidal spheres immersed in a solvent of 
depletion agent. The latter component can possess a polydisperse size
or shape distribution. Here we are mainly interested in mixtures where the
depletion potential between the spheres is strongly attractive already at
rather low concentrations of the depletion agent. In this case we can employ
the description of Asakura, Oosawa\cite{bib::asakura1954,bib::asakura1958}
and Vrij\cite{bib::vrij1976} for the mutual particle interactions, which
keeps the model simple. The colloid-colloid and the colloid-depletion agent
interactions are modeled as hard-core interactions, while the mutual
interaction between particles of the depletion agent vanish. It has been shown 
\cite{bib::schmidt2000} that the phase behavior of the Asakura-Oosawa-Vrij
model is closely related to the free-volume theory, in which the excess free
energy of the mixture is expanded up to first order in the density of the
depletion agent. This linearization of the excess free energy is equivalent to
a vanishing interaction among particles of the depletion agent. As
FMT\cite{bib::rosenfeld1989} provides a good framework for the free-volume
theory we first recall its main ideas as presented in
Refs.~\onlinecite{bib::dijkstra1999} and \onlinecite{bib::oversteegen2005} for
discrete mixtures and generalize it to continuous distributions of the depletion
agent.

\subsection{Free-volume theory}
 
In order to describe the phase behavior of colloidal mixtures one has first to
choose the thermodynamic ensemble. For our purpose a system with a fixed
number of big spherical colloidal particles $N_c$, which is coupled to a
reservoir of a depletion agent at chemical potential $\mu_d$ or fugacity $z_d$,
respectively, represents the best choice.\cite{bib::lekkerkerker1992} Hence,
the system is treated in a semi-grand canonical ensemble. The Helmholtz free
energy separates into two terms
\begin{eqnarray}
\beta F(N_c, V, z_d) &= &\beta F(N_c, V, z_d = 0) \nonumber\\&&+ \int_0^{z_d} \dd z_d' \left(
  \frac{\partial \beta F(N_c, V, z_d')}{\partial z_d'}\right).
\label{eqn::semiGCpotential}
\end{eqnarray}
The first term is the Helmholtz free energy of pure colloids. 
By Taylor expansion of the integrand of the second term in
Eq.~(\ref{eqn::semiGCpotential}) up to first order in $z_d$ we obtain an
expression for the volume available to particles of the depletion agent in  
a system of $N_c$ colloids at low concentrations of the depletion agent, the
so-called free volume denoted by $\alpha V$. In this limit of small fugacity
of the depletion agent, we can rewrite Eq.~(\ref{eqn::semiGCpotential})
into
\begin{equation}
\beta F(N_c, V, z_d) = \beta F_0(N_c, V) - \rho_d^r(z_d) \alpha V,
\label{eqn::freeEnergySemiGC}
\end{equation}
where $\rho_d^r$ is the reservoir density of the depletion agent. Within the
AOV model the free-volume fraction $\alpha$ depends only on the geometry of the
depletion agent, e.g. the size ratio $q=\sigma_d/\sigma_c$ in the case
of spherical polymer, and on the packing fraction of the colloidal
spheres $\eta_c = \pi \sigma_c^3 N_c/(6 V)$.

For the calculation of the phase diagram of the colloidal mixture we require
the chemical potential of the colloid component $\mu_c$ and the osmotic
pressure $p$. In the bulk two phases I and II coexist if they are in
chemical and mechanical equilibrium, i.e. if 
$\mu_c^{\textfoot{I}}=\mu_c^{\textfoot{II}}$ and
$p^{\textfoot{I}}=p^{\textfoot{II}}$, respectively. The chemical potential
$\mu_c$ and the pressure $p$ follow from the Helmholtz free energy
via\cite{bib::lekkerkerker1992} 
\begin{equation}
\mu_c = \frac{\partial F}{\partial N_c}\Bigg|_{V,z_d} = \mu_0(\eta_c) -
\rho_d^r(z_d) \left(\frac{\partial \alpha}{\partial \eta_c} \right) \frac{\pi
\sigma_c^3}{6} 
\label{eqn::chemicalpotential}
\end{equation}
and
\begin{equation}
p =  \frac{\partial F}{\partial V}\Bigg|_{N_c,z_d} = p_0(\eta_c) +
\rho_d^r(z_d) \left(\alpha - \eta_c \frac{\partial \alpha}{\partial \eta_c}
\right).   
\label{eqn::osmoticpressure}
\end{equation}
The chemical potential and the pressure of the pure reference system,
labeled $\mu_0$ and $p_0$, are described accurately by the Carnahan-Starling
expressions in the fluid phase. To assure thermodynamic consistency the
free-volume fraction $\alpha$ must also be calculated from the equation of
state of the reference system, i.e.~the Carnahan-Starling equation of
state.\cite{bib::oversteegen2005} Besides coexisting fluid states, i.e.
points on the binodal line in phase diagram, we calculate the spinodal line
which separates metastable states from thermodynamically unstable ones. For
thermodynamically unstable states the second derivative of the Helmholtz free
energy w.r.t.~the colloid packing fraction $\eta_c$ is negative and for points
on the spinodal line it vanishes, i.e. $\partial^2 F/\partial \eta_c^2 = 0$.

For the solid-fluid coexistence we need the chemical potential and equation of
state of a reference (fcc) hard-sphere crystal, which is described by
\cite{bib::frenkel1984}
\begin{equation}
\beta \mu = 2.1306 + 3 \ln{\left(\frac{\eta_c \eta_{cp}}{\eta_{cp}-\eta_c}\right)} 
+ \frac{3\eta_{cp}}{\eta_{cp}-\eta_c}
\end{equation}
with $\eta_{cp} = \pi\sqrt{2}/6\approx 0.74$, the packing fraction for close packing, and \cite{bib::wood1952}
\begin{equation}
\beta p v_s = \frac{3 \eta_c \eta_{cp}}{\eta_{cp}-\eta_c}.
\end{equation}

\subsection{Fundamental measure theory}

For discrete mixtures of two species the excess Helmholtz free energy density
$\Phi = \beta F / V$ is constructed from two sets of weighted densities
$\{n_i^c\}$ and $\{n_i^d\}$, one for the colloids $c$ and one for the depletion
agent $d$.\cite{bib::schmidt2000} In homogeneous bulk fluids $i$ labels four
scalar-weighted densities, which can be identified as the scaled-particle
theory variables $\zeta_i$ of the $N$-component
mixture\cite{Rosenfeld94,Rosenfeld95}
\begin{eqnarray}
n_i \equiv \zeta_i = \sum_{\nu=1}^{N} g_{\nu}^{(i)} \rho_{\nu},
\end{eqnarray}
with the geometrical measures of component $\nu$: volume $g_{\nu}^{(3)}
\equiv v_{\nu}$, surface area $g_{\nu}^{(2)} \equiv a_{\nu}$, integrated 
mean curvature $g_{\nu}^{(1)} \equiv c_{\nu}$ and Euler
characteristics $g_{\nu}^{(0)} \equiv X_{\nu}$.

An explicit expression for the free energy density of the AOV mixture in
the limit of low density of the depletion agent can be obtained by
expanding the hard-sphere (HS) mixture free energy with weighted densities
$n_i = n_i^c + n_i^d$ up to linear order in the density
$\rho_d$:\cite{bib::schmidt2000,bib::oversteegen2005}
\begin{eqnarray}
\Phi^{\textfoot{AOV}} (\{n_i^c\},\{n_i^d\}) & = & \Phi^{\textfoot{HS}}(\{n_i^c\}) +
\sum_{k=0}^{3} \frac{\partial \Phi (\{n_i^c\})}{\partial n_k^c} n_k^d .
\label{eqn::phiAO}
\end{eqnarray}
In order to relate the terms in Eq.~(\ref{eqn::phiAO}) to the free-volume
fraction $\alpha$ we transform the excess Helmholtz free energy density, which
in Eq.~(\ref{eqn::semiGCpotential}) is given in the semi-grand canonical
ensemble, into the canonical ensemble and obtain in terms of the colloid
packing fraction $\eta_c$ and the density $\rho_d$ 
\begin{equation}
\Phi(\eta_c, \rho_d; q) = \Phi (\eta_c) - \rho_d \ln{( \alpha(\eta_c; q) )}.
\label{eqn::phiSemiGrandCan}
\end{equation}
By direct comparison of Eq.~(\ref{eqn::phiAO}) and
Eq.~(\ref{eqn::phiSemiGrandCan}) we obtain a FMT expression for the
free-volume fraction\cite{bib::oversteegen2005}
\begin{equation}
\alpha(\eta_c; \{g_{\nu}^{(i)}\}) = \exp{\left( - \sum_{k=0}^3 \frac{\partial
      \Phi(\{n_i^c\})}{\partial n_k^c}~ g_d^{(k)} \right)}.
\end{equation}
The partial derivatives of the free energy w.r.t.~the weighted densities
are thermodynamic quantities of the (pure) colloids, namely, the pressure
$p_0$, the surface tension $\gamma$ at a planar hard wall and the bending
rigidities $\kappa$ and $\bar\kappa$, which describe the effect on the free
energy due to curved
surfaces. \cite{koenig2004,bib::oversteegen2005,bib::hansengoos06a} The
depletion agent enters 
the expression for the free-volume fraction $\alpha$ only through its
geometrical measures $g_d^{(k)}$. For the free energy density we employ
Rosenfeld's original formulation of FMT\cite{bib::rosenfeld1989} and the more
accurate White Bear version from Refs.~\cite{bib::roth2002,bib::yu2002}
\begin{eqnarray}
\Phi(\{ n_{\alpha} \})&=& - n_0 \ln(1-n_3) + \frac{n_1 n_2}{1 - n_3}
\nonumber \\ &&+\frac{n_2^3 \left(n_3 + (1-n_3)^2 \ln{(1-n_3)}\right) }{36 \pi n_3^2
  (1-n_3)^2} , 
\label{eqn::phiWhitebear}
\end{eqnarray}
which is based on the Boublik-Mansoori-Carnahan-Starling-Leland (BMCSL)
equation of state\cite{bib::boublik1970,bib::mansoori1971}
\begin{eqnarray}
\beta p_{BMCSL} &=& \frac{n_0}{1-n_3} + \frac{n_1 n_2}{(1-n_3)^2} +
\frac{n_2^3}{12 \pi (1-n_3)^3} \nonumber\\ &&- \frac{n_3 n_2}{36 \pi (1-n_3)^3}.
\label{eqn::pressureMCSL}
\end{eqnarray}
It would also be possible to base the free-volume theory on a recent, slightly
more consistent, mixture equation of
state.\cite{bib::hansengoos06a,bib::hansengoos06b}

\subsection{Generalization to polydisperse depletion agent}
FMT is a mixture theory by construction. For mixtures
of colloids and ideal depletion agent we can employ parameters $q_i$ to
parameterize the geometry of particles of the depletion agent components.
Although it is straightforward within FMT
to treat general mixtures of $\nu_c$ colloid and $\nu_d$ depletion agent
components, we will in the following restrict our studies to the simpler case
of a single colloid species with packing fraction $\eta_c$. For the depletion
agent we start by considering $\nu_d$ components. In the semi-grand canonical
ensemble the excess Helmholtz free energy density is, 
analogous to Eq.~(\ref{eqn::phiSemiGrandCan}), given by
\begin{equation}
\Phi(\eta_c, \{\eta_{d,i}^r\}; \{q_i\}) = \Phi_c - \sum_{\nu=1}^{\nu_d}
\rho_{d,i}^r \alpha(\eta_c; \{q_i\}), 
\label{eqn::phiSemiGrandCan2}
\end{equation}
and the set $\{g_{\nu}^{(i)}\}$ of geometrical measures is characterized by
the set of parameters $\{q_i\}$. To study the influence of polydispersity in
the depletion agent on the phase behavior it might be sufficient to consider a
discrete mixture, however we prefer to introduce continuous distributions,
which can represent experimental systems more accurately than a discrete
distribution. In the following we restrict the set $\{g_{\nu}^{(i)}\}$ to
depend on a single parameter $q$ only. We introduce $\distvar(q)$ as a
continuous distribution of the depletion agent in the reservoir, and require
that $\int \distvar(q) \dd q = 1$. The density distribution of the depletion
agent follows directly: $\rho_d^r(q) = \rho_d^r \, \distvar(q)$. By specifying the
fundamental geometrical measures of the depletion agent as function of $q$ we
obtain, similar to Ref.~\onlinecite{bib::warren1997}, 
\begin{eqnarray}
&&\Phi^{pd} (\eta_c, \rho_d^r; [\distvar]) = \Phi(\eta_c) \nonumber\\&&- \rho_d^r \int  \dd q \, \distvar(q) \alpha(\eta_c, v_d(q), a_d(q), c_d(q), X_d(q) ),
\label{eqn::phiPolydisperse}
\end{eqnarray}
which is a functional of the distribution $d(q)$. Note that the precise
meaning of $q$ is unspecified so far and can refer to a size ratio, in the
case of spherical mixtures, or to a parameter that specifies the shape of the
depletion agent in a more complicated manner. We return to this point
later. Analogous to Eq.~(\ref{eqn::freeEnergySemiGC}) we call the integral
on the r.h.s.~of Eq.~(\ref{eqn::phiPolydisperse}) the effective free-volume
fraction $\alpha_{\textrm{eff}}$ and rewrite Eq.~(\ref{eqn::phiPolydisperse})
as 
\begin{equation}
\Phi^{pd} (\eta_c, \rho_d^r; [\distvar]) = \Phi(\eta_c) - \rho_d^r
\alpha_{\textrm{eff}}(\eta_c; [\distvar]). 
\label{eqn::phiPolydisperseAlphaEff}
\end{equation}

%
\section{Distribution Functions}\label{sec::distributions}

%
To study the phase behavior of a mixture of spherical colloids and a
polydisperse depletion agent we focus on three frequently used
distributions, namely the Schulz ($S$), Gaussian ($G$) and the Hat-like
($H$) distributions. These distributions are characterized by two parameters
corresponding to its first and second moment. The average asymmetry ratio is
given by $\bar q = \tilde \sigma_d/\sigma_c$, with the colloid diameter
$\sigma_c$ and a length-scale of the depletion agent $\tilde \sigma_d$. The
parameter $z$ describes 
the degree of polydispersity: the limiting case $z\to \infty$ equals the
monodisperse fluid and by decreasing the value of $z$ the distribution is
broadened. The distributions under consideration here lead to explicit
expressions for the effective free-volume fraction
$\alpha_{\textrm{eff}}(\eta_c; \distvar(q;\bar q, z))$, if the geometries of
the depletion agent is simple. For more complicated shapes we have to perform
the integration numerically. In terms of $\bar q$ and $z$ the distributions
considered here are given by
\begin{eqnarray}
\distvar_{S}(q;\bar q, z)  & = & \left(\frac{z}{\bar q}\right)^z  q^{z-1}
\frac{\exp{(-z q/ \bar q)}}{\Gamma(z)}, \quad z \geq 1, 
\label{eqn::schulzDist} \\
\distvar_{G}(q;\bar q, z)  & = & \frac{z}{\sqrt{\pi}}  \exp{( - (q-\bar q)^2
z^2)}\nonumber\\
                          &&     \times        \, 2 \, \Theta (q) / [\erf(z \bar q)+1],
\label{eqn::gaussianDist}\\
\distvar_{H}(q;\bar q, z)  & = & \frac{z}{2} \Theta\left(q-\bar
q+z^{-1}\right)\nonumber\\
&&\times\Theta\left(-q+\bar q+z^{-1}\right), \quad \quad \bar q z > 1. 
\label{eqn::hatlikeDist}
\end{eqnarray}
It is evident that $q\!>\!0$ is required in all distributions and the
integration in $\alpha_{\textrm{eff}}(\eta_c; \distvar(q;;\bar q, z))$ is
performed in the limits from 0 to $\infty$. In the case of a Gaussian
distribution one can obtain simpler explicit expressions by imposing the full
integration range $q=-\infty$ to $\infty$. The result, $\distvar'_{G}$, is
similar to that of the cut Gaussian distribution $\distvar_{G}$
\begin{equation}
\distvar'_{G}(q;\bar q, z) =  \frac{z}{\sqrt{\pi}}  \exp{( - ( q - \bar q )^2
  z^2 )}.                       
\label{eqn::extGaussianDist}
\end{equation}

Obviously, it is possible to consider different size distributions such as the
log-normal distribution, which decays slower for large values of $q$ than the
distributions considered here.

%
\subsection{Explicit expressions for $\alpha_{\textrm{eff}}$} 

The equations for the chemical potential $\mu_c$ and the osmotic pressure $p$
from Eqs.~(\ref{eqn::chemicalpotential}) and (\ref{eqn::osmoticpressure}) for
polydisperse size distributions contain integrals over the distribution.
For the calculation of the phase diagram these equations
must be solved simultaneously. In general this must be done numerically.
In the case of one- and two-dimensional depletion agents, such as infinitely
thin rods or platelets, their geometry is sufficiently simple so that the
effective free-volume fraction $\alpha_{\textrm{eff}}$ can be calculated
explicitly. Infinitely thin needles represent an one-dimensional depletion agent.
Infinitely thin platelets represent a two-dimensional depletion agent. Here we
consider both disk- and hexagon-shaped platelets. The geometric measures for
these geometries are given in Tab.~\ref{tab::geometricMeasures}. In
Appendix~\ref{app::explicitExpressions} we show explicit expressions for
$\alpha_{\textrm{eff}}$ for one- and two-dimensional depletion agents and the
distributions given above. As aforementioned, we obtain the
thermodynamic quantities of the pure hard-sphere fluid that enter the
free-volume fraction $\alpha_{\textrm{eff}}$ from the excess free energy
density $\Phi$ of the White Bear version of
FMT.\cite{bib::roth2002,bib::yu2002} These quantities are the pressure $\beta
p=\partial \Phi/\partial n_3$, the surface tension $\beta \gamma = \partial
\Phi / \partial n_2$ and bending rigidities $\beta \kappa = \partial \Phi /
\partial n_1$ and $\beta \bar\kappa = \partial \Phi / \partial
n_0$.\cite{koenig2004,bib::oversteegen2005,bib::hansengoos06a}

In the present study we restrict the product $\bar q z$ to sufficiently large
values. For the Hat-like distribution we require $\bar q z>1$ which ensures
that $\int_0^{\infty}\dd q \, \distvar(q) = 1$. In the case of the Gaussian
distribution $\distvar'_{G}$ we require $\bar q z>1$ because for small values 
of $\bar q z$ the contribution of the distribution from values $q\!<\!0$ increases,
e.g.~for $\bar q=0.5$ and $z=3$ we obtain $\int_{-\infty}^{0}\dd q \,
\distvar_G(q) \sim {\mathcal O}(10^{-2})$.

In the case of two-dimensional depletion agents, the geometric measures can
differ for different geometry of the depletion agent: for disk- and hexagonal
platelets $a_d$ and $c_d$ differ -- see Tab.~\ref{tab::geometricMeasures}.
For that reason we show the result for $\alpha_{\textrm{eff}}$ in
Appendix~\ref{app::explicitExpressions} in its general form.

\begin{table}
\begin{tabular}{cl|l|cccc}
\hline
Dim       & Depletion agent     &  Geometry      &  $v_d$  &  $a_d$  &  $c_d$  &  $X_d$ \\
\hline
\hline
1         & needle    & $L=q\, \sigma_c$,  $R = 0 $  & 0 & 0 & $\frac{L}{4}$ & 1 \\
\hline
2         & disk & $R=q\, \sigma_c/2$, $h=0$      & 0 & $2 \pi R^2$    & $\frac{\pi}{4}R$ & 1 \\
2         & hexagon &  $R=q\, \sigma_c/2$,  $h=0$ & 0 & $3\sqrt{3}R^2$ & $\frac{3}{4}R$ & 1 \\     
\hline
\end{tabular}
\caption{Geometrical measures for one- and two-dimensional depletion
  agents. All lengths are measured in units of the diameter $\sigma_c$ of the
  spherical colloids. We consider needles of length $L$ and radius $R=0$ and
  platelets with radius $R$ and thickness $h=0$. The value of $q$ plays
  the role of a size ratio and compares the size of the depletion agent to
  that of a colloid.}   
\label{tab::geometricMeasures}
\end{table}

%
\section{Results} \label{sec::results}

While our approach developed in Section~\ref{sec::theory} is general and
allows one to study various systems, we want to present only one
application. We study the phase behavior of a ternary mixture of
(monodisperse) colloidal spheres with two polydisperse polymer components of
distinct sizes. Both polymer components possess a size distribution which we
describe by Schulz distributions, which in total lead to 
$\distvar_{S}(q;\bar q, \bar Q, z_{\bar q}, z_{\bar Q}) = x \,
\distvar_{S}(q;\bar q, z_{\bar q}) + (1-x) \, \distvar_{S}(q;\bar Q, z_{\bar
  Q})$ with a mixing parameter $x$ and two averaged size ratios $\bar q$ and
$\bar Q$. For simplicity we assume the width of both parts of the distribution
to be equal. The polymer reservoir packing fraction is given by $\tilde
\eta_d^r = \frac{\pi}{6} \langle q \rangle^3 \sigma_c^3 \rho_d^r$, where
$\langle q \rangle = x \, \bar q + (1-x) \, \bar Q$.

We expect the phase behavior to be most interesting if the average size ratios
$\bar q$ and $\bar Q$ are distinct, so that the phase diagrams in the limiting
cases $x=0$ and $x=1$ are sufficiently different. To this end we choose $\bar
q=0.25$ and $\bar Q=2.0$. In the case of $x=0$ when we obtain a binary mixture
with an averaged size ratio $\bar Q=2.0$, the phase diagram exhibits a stable
fluid-fluid phase separation into a colloid poor gas and a colloid rich liquid
phase. For the mixing parameter $x=1$, when the averaged size ratio 
$\bar q=0.25$, the fluid-fluid phase separation, which is metastable w.r.t.~to
crystallization in the monodisperse case,\cite{bib::dijkstra1999} can be
stabilized by a sufficient degree of
polydispersity.\cite{bib::meijer1994,bib::sear1997,bib::fasolo2005}

Before we study the full phase diagram of the ternary mixture, we determine
which value of $z_{\bar q}$ is required in order to stabilize the fluid-fluid
phase separation for $x=1$.

\subsection{Influence of size polydispersity} \label{sec::sizepoly}

In this section we study the phase diagram of a binary colloid-polymer
mixture, which corresponds to the limiting case $x=1$, with an averaged size
ratio of $\bar q=0.25$. In contrast to the aforementioned examples of one-
and two-dimensional depletion agents, $\alpha_{\textrm{eff}}$ can not be
obtained explicitly. Therefore, we have to determine the phase coexistence by
solving two equations, Eqs.~(\ref{eqn::chemicalpotential}) and
(\ref{eqn::osmoticpressure}), numerically. 

The equation of state of the pure reference system and the free-volume 
fraction $\alpha_{\textrm{eff}}$ respectively, and hence the chemical
potential $\mu_c$ and the osmotic pressure $p$ of the colloids, are based on
the free energy density $\Phi$ of the White Bear version of FMT,
Eq.~(\ref{eqn::phiWhitebear}). 

\begin{figure}[t]
\centering\epsfig{file=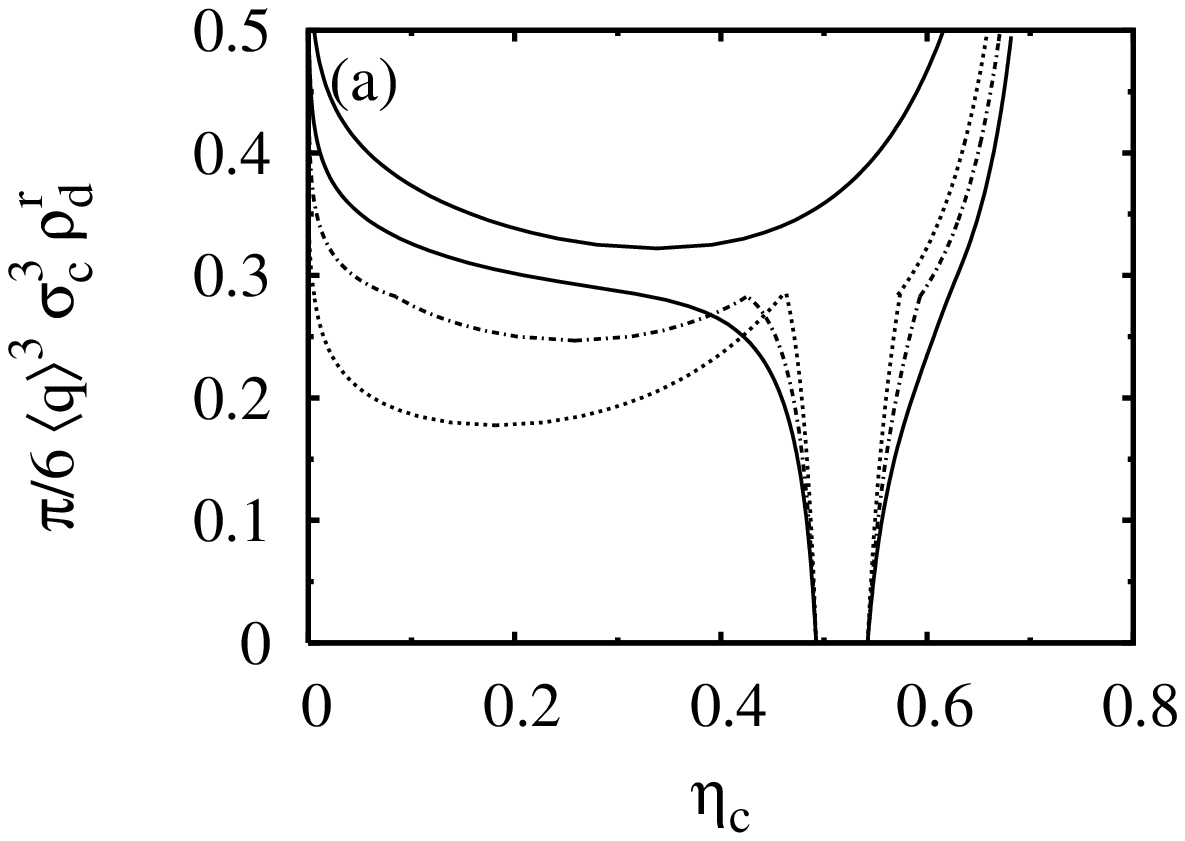,width=0.5\textwidth, angle=0}
\centering\epsfig{file=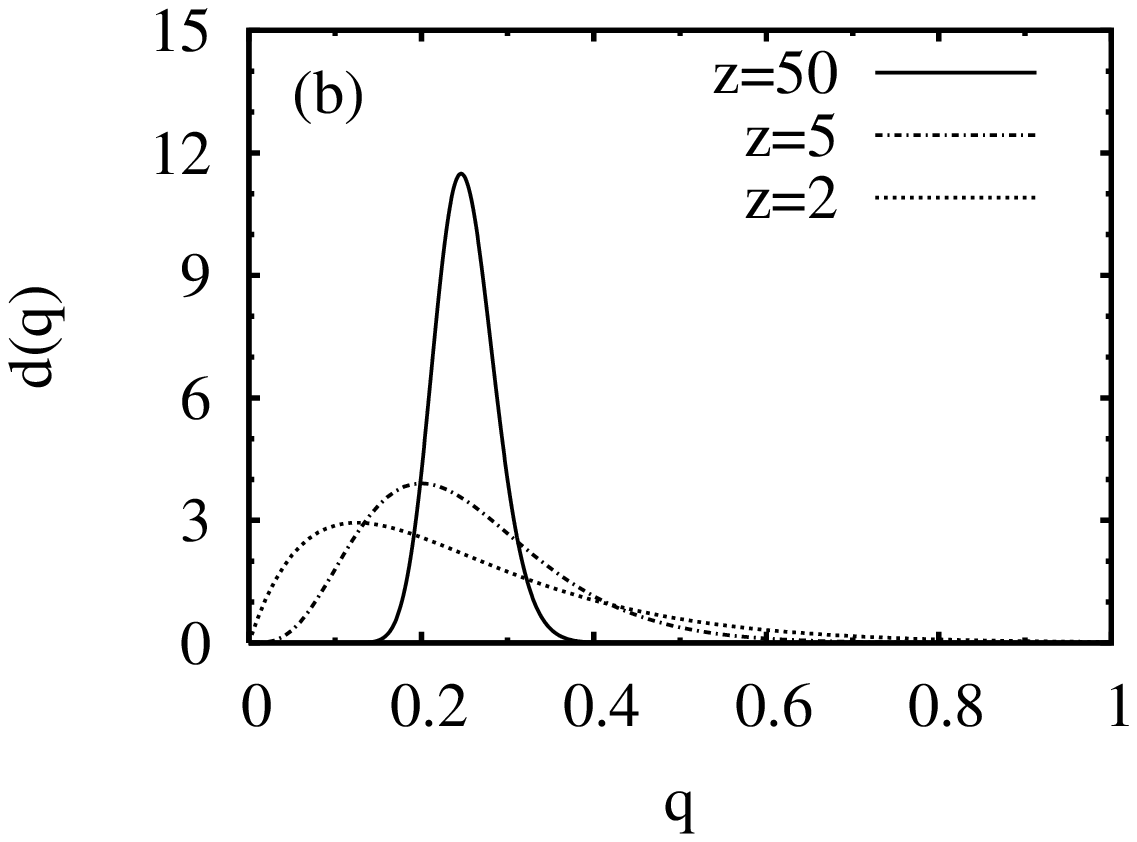,width=0.5\textwidth, angle=0}
\caption{\label{fig::phasediagramPolyAO} (a) Phase diagrams of mixtures of
  colloids and polydisperse (spherical) polymer with an average size ratio
  $\bar q=0.25$ and (b) the corresponding distributions. We employ a
  free-volume theory based on the BMCSL equation of state. For low degrees of
  polydispersity, corresponding to $z=50$ (full lines), the fluid-fluid phase
  separation is metastable w.r.t.~fluid-solid coexistence. Upon increasing the
  degree of polydispersity, corresponding to $z=5$ (dashed-dotted lines) and
  $z=2$ (dotted lines), the fluid-fluid coexistence is stabilized.}
\end{figure}

In Fig.~\ref{fig::phasediagramPolyAO}(a) we show the phase diagrams of
polydisperse AOV mixtures as we vary the degree of polydispersity. 
For the size ratio $\bar q=0.25$ and a narrow distribution of the polymer, the
fluid-fluid phase separation remains metastable similar to the monodisperse
case. We show the phase diagram for $z=50$ (full lines) in
Fig.~\ref{fig::phasediagramPolyAO}(a) and the corresponding size distribution
in Fig.~\ref{fig::phasediagramPolyAO}(b). Upon decreasing $z$, which
broadens the size distribution, the fluid-fluid and the fluid-solid
coexistence lines come closer to each other and almost touch for $z=10$ (not
shown). By further increasing the degree of polydispersity the fluid-fluid
coexistence becomes stable, as we show in
Fig.~\ref{fig::phasediagramPolyAO}(a) for $z=5$ (dashed-dotted lines) and
$z=2$ (dotted lines). The corresponding size distributions are shown in 
Fig.~\ref{fig::phasediagramPolyAO}(b). The critical point is shifted to lower
colloid packing fractions $\eta_c$, as the value of $z$ decreases. 

The polymer distribution in the system differs from that in the reservoir,
\cite{bib::sear1997} which is a phenomenon known as fractionation. For
spherical polymer we find results equivalent to those of
Ref.~\onlinecite{bib::sear1997}. At low packing fractions of the colloids the
size distribution of (spherical) polymer follows closely the reservoir
distribution. As the colloid packing fraction increases, the maximum of the
polymer size distribution decreases significantly and moves towards smaller
polymer radii. In principle we observe a qualitatively similar but less
pronounced behavior also in the case of platelet-like or rod-like depletion
agents. The maximum in the size distribution of the depletion agent becomes
smaller and moves to smaller sizes as the colloid packing fraction in the
system increases. This is to be expected, because one- or two-dimensional
depletion agents require less free volume to fit between colloids than
spherical polymer.

We have confirmed that a treatment of the phase behavior completely based on 
the PY compressibility equation of state yield qualitatively similar
results. In the present approach this is achieved by employing the free energy
density $\Phi$ of Rosenfeld's original formulation of FMT to evaluate
$\alpha_{\textrm{eff}}$ and the pressure and chemical potential for the pure
reference system. The results in Refs.~\onlinecite{bib::meijer1994}, 
\onlinecite{bib::sear1997}, and \onlinecite{bib::fasolo2005} are somewhat
in between these two predictions as their approach employs a hard-sphere
reference system based on the Carnahan-Starling equation of state and a
free-volume fraction based on the PY compressibility equation of state, which
for spheres is equivalent to the scaled-particle theory expression.

%
\subsection{Influence of shape polydispersity}
  \label{sec::shapepoly} 

In this section we briefly address the question whether polydispersity,
which describes a shape distribution, can also stabilize the fluid-fluid phase
separation. One can think of the shape polydispersity in two different
ways. On the one hand it can be thought as describing effects due to
imperfections in the synthesis of colloidal particles. On the other hand, and
this is more relevant in the present context, it can be seen as a simple model
for the effect that polymer coils are not rigid and can change their average
shape when squeezed in between colloids.

\begin{figure}[t]
\centering\epsfig{file=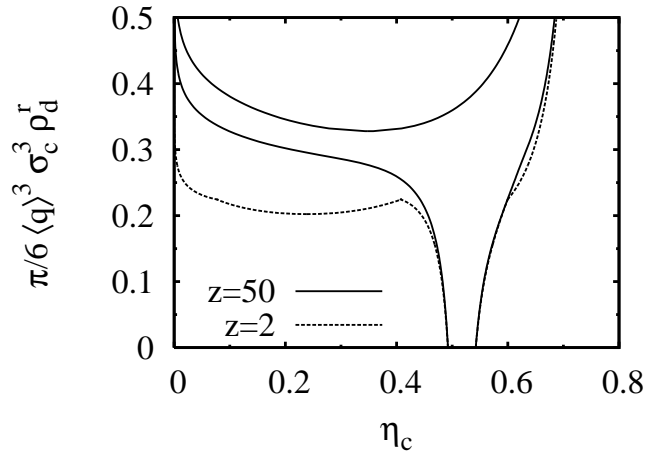,width=0.5\textwidth, angle=0}
\caption{\label{fig::shape} Effect of shape polydispersity of the
 depletion agent on the phase behavior of a mixture of colloids and depletion
  agent. The depletion agent consists of ellipsoids with half-axes $a=b=\sigma_d/(2
  \sqrt{q})$ and $c=q \sigma_d/2$. The mean value of the distribution is fixed
  at $\bar q=1.0$ while the size ratio between the depletion agent and the colloid
  is fixed at $\sigma_d/\sigma_c=0.25$. For $z=50$ (full lines) the fluid-fluid
  coexistence is metastable
  and similar to the monodisperse case (not shown here). Upon increasing the 
  degree of polydispersity we find two stable fluid phases. The phase diagram
  for $z=2$ (dashed lines) exhibits a stable critical point and a triple point.}
\end{figure}

To this end we consider ellipsoids, with half-axes $a=b=\sigma_d/(2 \sqrt{q})$
and $c=q \sigma_d/2$, for which the geometrical measures can be obtained
explicitly.\cite{bib::oversteegen2005} This case is relatively simple as a
single parameter $q$ characterizes the shape while $\sigma_d=0.25 \sigma_c$ is fixed. 
Obviously, more complicated
scenarios are possible. Note that for this choice of half-axes the volume $v_d
= \pi \sigma_d^3 /6$ is kept fixed, while the surface area $a_d=\pi \sigma_d^2
(1/q + q^{3/2} (\arccosh{1/q^{3/2}})/\sqrt{1/q-q^2}) / 2$ and integrated mean
curvature $c_d=\sigma_d (q + (\arccos{q^{3/2}})/\sqrt{q-q^4}) / 4$ depend on
$q$. The Euler characteristic $X_d=1$, independent of $q$. Here, the parameter
$q$ describes the degree of deviation of the shape from a sphere: for small
values of $q$ we obtain lens-shaped particles (oblates), while for
large values of $q$ the particles become cigar-shaped (prolates). Similarly to
the case of size polydispersity, we observe that an increasing degree of shape
polydispersity stabilizes the fluid-fluid phase separation w.r.t.~fluid-solid
decomposition. In Fig.~\ref{fig::shape} we show an example for a narrow
shape distribution with $z=50$ (full line), in which case the fluid-fluid
critical point is metastable, and one for a broad distribution with $z=2$
(dotted line). For $z=2$ we find a stable fluid-fluid phase separation.

It seems that polydispersity in general favors a stable fluid-fluid phase
separation because the polydisperse depletion agent can fill the free volume
of the system more effectively than a monodisperse one. Even if the bigger
particles of the distribution cannot find free volume in the system, the
smaller ones still do. In terms of the depletion potential this would result
in a longer range of the effective attraction between
colloids,\cite{bib::goulding2001} giving rise to a more negative effective
second virial coefficient $B_2^{\textrm{eff}}$ which increases the tendency
for fluid-fluid phase separation.\cite{bib::roth2001} This effect seems to be
robust against details of introducing polydispersity and can also be observed
for a depletion agent with $a=b=\sigma_d/2$ and $c=q \sigma_d/2$, where all 
geometrical measures, except of the Euler characteristics, depend on $q$.

%
\subsection{Ternary mixtures with a bimodal polydisperse distribution}
  \label{sec::bimodal} 

Based on our findings in Section~\ref{sec::sizepoly} we choose 
$z_{\bar q}=z_{\bar Q}=5$ for the bimodal polydisperse polymer distribution. 

\begin{figure}[t]
\centering\epsfig{file=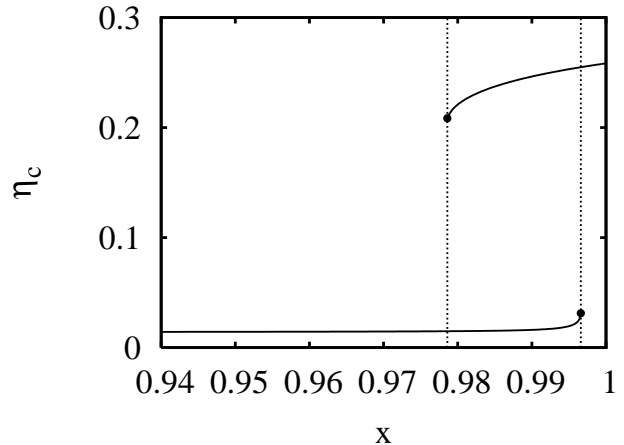,width=0.5\textwidth, angle=0}
\caption{\label{fig::criticalpoints}
  The path of the two critical points in a ternary mixture of colloidal spheres
  and two polydisperse polymer components. If the mixing parameter $x=1$,
  only the smaller polymer are present and we find a stable critical point at
  $\eta_c=0.258$ and $\eta_d^r=0.247$. As we decrease the value of $x$ this
  critical point moves towards smaller values of $\eta_c$ and vanishes at
  $x\approx 0.979$ (upper line). For a mixing parameter $x=0$, when only the
  bigger polymer are present, we find a second critical point at
  $\eta_c=0.014$ and $\eta_p^r=0.924$, which moves towards larger values of
  $\eta_c$ (lower line). For  $x \in [0.979,0.997]$ (indicated by the dotted
  lines) we find two critical points in the system.}
\end{figure}

Before we show the calculated phase diagrams, it is interesting to compare the
two {\em stable} critical points corresponding to the limiting cases,
i.e. $x=1$ and $x=0$. The critical points can be found at very distinct
regions of the phase diagram. In Fig.~\ref{fig::criticalpoints} we show the
paths of the critical points through the phase diagram as a function of the
mixing parameter $x$. When we start at $x=1$, for which only the smaller
polymer are present, we find a critical point at $\eta_c=0.258$ and
$\eta_d^r=0.247$. As we decrease the value of $x$, the critical point moves
towards smaller values of $\eta_c$ until this critical point becomes
metastable. For $x\lesssim 0.979$ the critical point no longer exists. When we
start at $x=0$, for which only the larger polymer are present, we find a
second critical point at $\eta_c=0.014$ and $\eta_d^r=0.924$. As the value of
$x$ is increased we find that this critical point moves towards larger values
of $\eta_c$ until it becomes metastable. For $x\gtrsim 0.997$ this critical
point vanishes.

It is most interesting to note that both critical points found in this
system, which correspond to critical points of the limiting cases, can exist
at the same time for a range of values of the mixing parameter, $x \in
[0.979,0.997]$. In order to obtain a full picture and determine where 
these critical points are stable we calculate the full phase
diagrams in this regime. 

\begin{figure*}[ht]
\includegraphics[width=0.5\textwidth,angle=0]{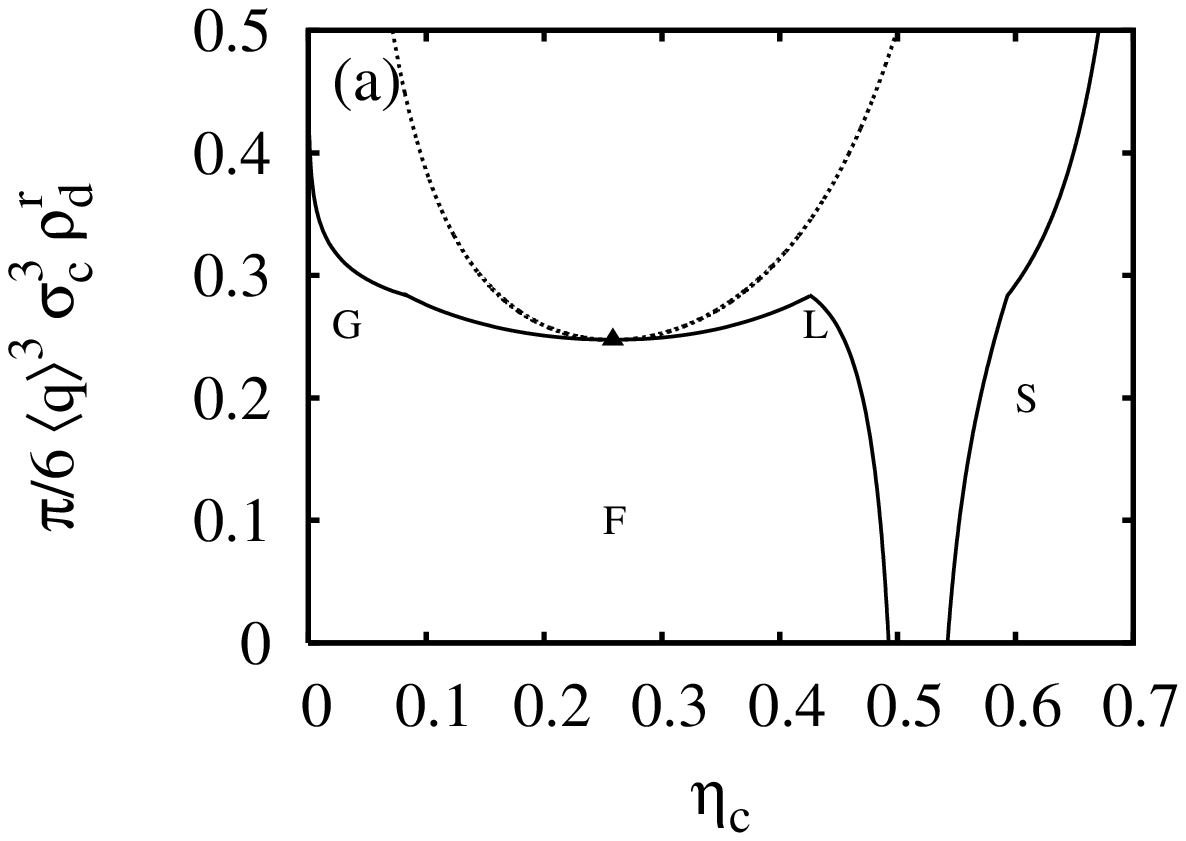}~\includegraphics[width=0.5\textwidth,angle=0]{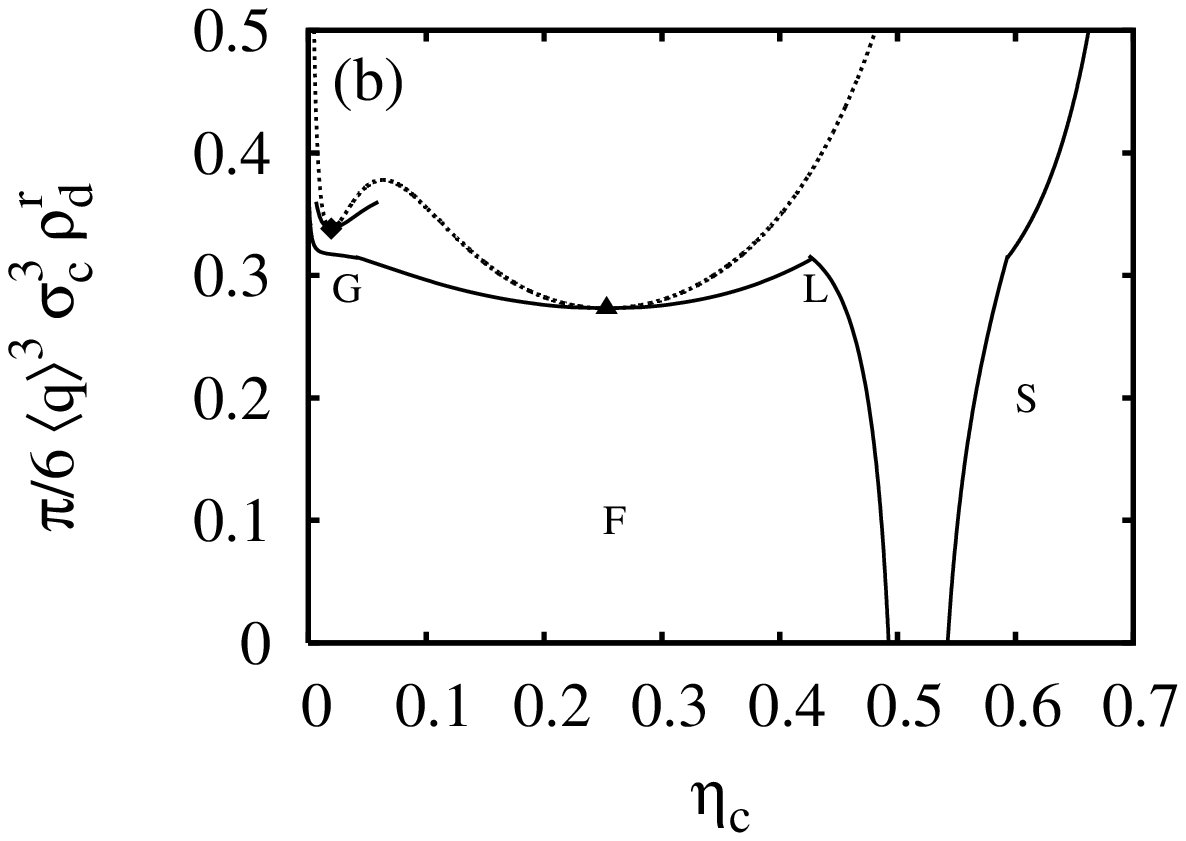}
\includegraphics[width=0.5\textwidth,angle=0]{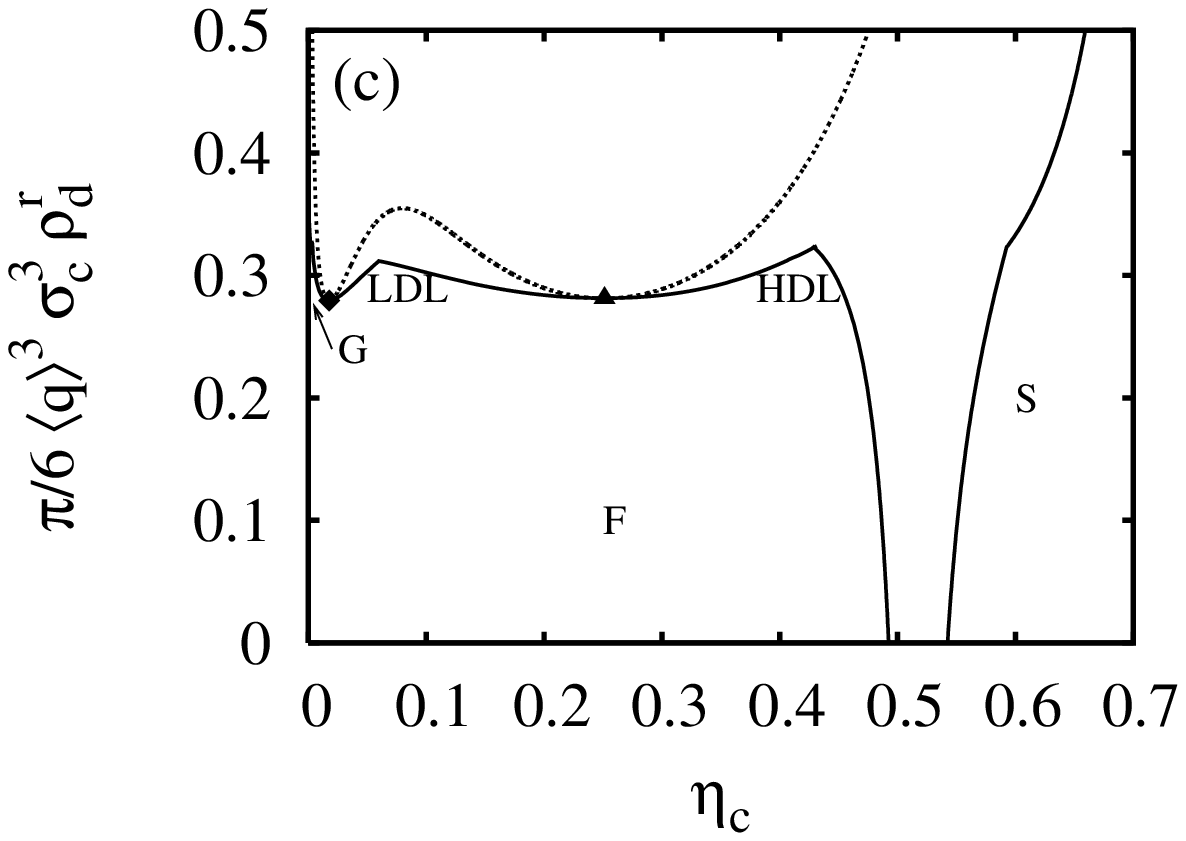}~\includegraphics[width=0.5\textwidth,angle=0]{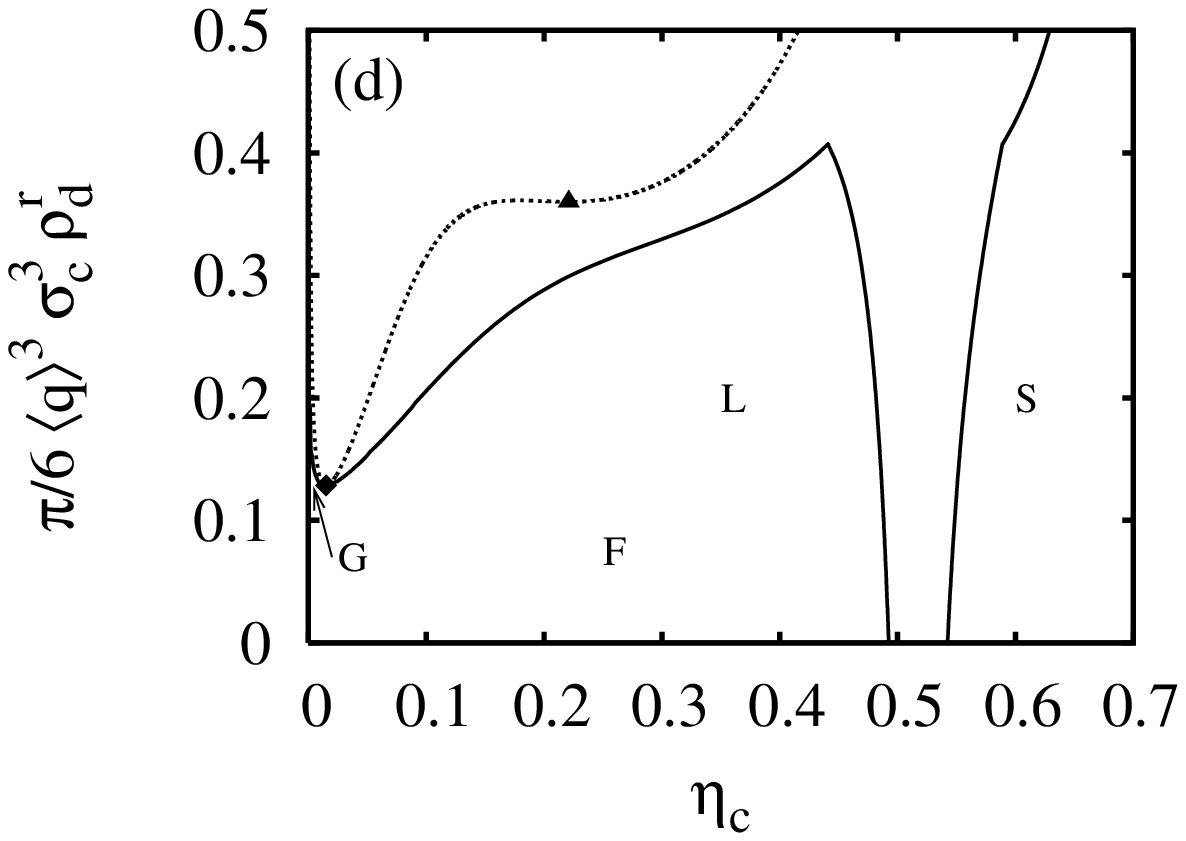}
\caption{\label{fig::ternaryPhaseDiagrams}
  Phase diagrams for a mixture of colloids and a bimodal distribution of
  polydisperse (spherical) polymer. For $x$ very close to 1 we recover the
  phase diagram for the monomodal distribution, see plot (a) for
  $x=0.9999$. The fluid-fluid coexistence (full line) is stabilized by
  polydispersity. Upon decreasing $x$, i.e. by adding larger polymer, the
  spinodal (dotted line) exhibits a second minimum for $x=0.995$ in Fig. (b)
  and we observe a second critical point, which is metastable (diamond). For
  $x$=0.9935 we observe two stable critical points. Beside a low density (in
  colloids) gas (G) there exists a low density (LDL) and a high density (HDL)
  liquid phase -- see Fig. (c). For a value of $x=0.98$, shown in (d), the
  critical point at high values of $\eta_c$ (triangle) is metastable.}
\end{figure*}

In Figs.~\ref{fig::ternaryPhaseDiagrams}(a)-(b) we show phase diagrams
in the $\eta_c$-$\tilde\eta_d^r$--representation for different values of
$x$. For $x=0.9999$ [see Fig.~\ref{fig::ternaryPhaseDiagrams}(a)] the phase
diagram is similar to the monodisperse case studied above. Upon decreasing
$x$, the fraction of polymer with average size ratio $\bar Q$ increases, and
the spinodal displays two minima, one corresponding to the stable critical
point (triangle) and the second minimum corresponding to an additional
metastable fluid-fluid phase separation (diamond) -- see
Fig.~\ref{fig::ternaryPhaseDiagrams}(b). Further decreasing the value of the 
mixing parameter to $x=0.9935$ [Fig.~\ref{fig::ternaryPhaseDiagrams}(c)]
yield a phase diagram with two {\em stable} critical points and a second
triple point. At one triple point, a gas, a liquid, and a solid phase coexist,
while at the second triple point a gas phase and two liquid phases with
different densities coexist. There is a novel low density liquid phase in the
region between the two critical points with colloid  packing fractions
$\eta_c=0.0177$ and $0.251$. As we vary the value of $x$ the triple points
move. According to Gibbs' phase rule a four-phase coexistence between a gas,
the low density liquid, the high density liquid and the solid is also
possible. For our system it would be expected for $x$ between the values used
in Fig.~\ref{fig::ternaryPhaseDiagrams}(b) and (c). 
In Fig.~\ref{fig::ternaryPhaseDiagrams}(d) for
$x=0.98$ we also observe two critical points. The critical point with lower
colloid packing fraction (diamond) is stable. The second critical point at
higher colloid packing fraction (triangle) is now in the metastable regime and
vanishes upon decreasing the values of $x$. For these systems one liquid phase
and one triple point are stable. Further decreasing the values of $x$ 
below approximately 0.8 lead to a shift of the coexistence region 
to higher values of $\tilde \eta_d^r$. For $x\ll 1$ the fluid-fluid binodal 
is in the region of the fluid with monomodal polydisperse depletion agent 
with $\bar Q=2.0$.

We observe similar behavior for a mixture of colloids and two monodisperse
polymer components, i.e. $z_{q}$ and $z_{Q} \to \infty$. For the same size
asymmetries as above, $\bar q \to q=0.25$ and $\bar Q \to Q=2.0$,
the critical point is metastable when $x=1$, or equivalently $\langle q
\rangle = 0.25$. We observe one stable and one metastable critical point 
for low colloid packing fraction upon decreasing the value of $x$. 
In the case $q=0.5$ and $Q=2.0$ there is a stable critical point for
$x=1$. Decreasing $x$, similarly to the polydisperse case described above,
leads to a second stable critical point. To summarize, we find that both by
means of choosing the degree of polydispersity and the size asymmetry we can
treat a model fluid which exhibits one or two (stable) critical points and one
or two (stable) liquid phases. Furthermore, we have verified that it is
possible to generate two critical points and hence {\em three} fluid phases by
using a depletion agent of different than spherical shape. 

%
\section{Discussion}\label{sec::discussion}

We have generalized a novel approach to free-volume
theory\cite{bib::oversteegen2005} based on
FMT\cite{bib::rosenfeld1989,Rosenfeld94,Rosenfeld95} to treat a depletion agent 
of arbitrarily shaped particles with a continuous distribution. Here, we
introduced one parameter, $q$, to specify a continuously varying distribution
of depletion agent.

To evaluate the phase diagram within FVT an expression for the free-volume
fraction is required. In the case of polydisperse size or shape distributions
for the depletion agent, one can obtain an {\em effective} free-volume
fraction, which in general has to be calculated numerically. In the cases
of infinitely thin needles or platelets it is possible to get explicit
expressions for the effective free-volume fraction for various distributions
which are of interest from an experimental and a theoretical point of view. We
give some expressions for $\alpha_{\textrm{eff}}$ in
Appendix~\ref{app::explicitExpressions}. 

Beside size polydispersity, we have also considered {\em shape}
polydispersity. To this end we have calculated phase diagrams of a mixture of
colloidal spheres with penetrable ellipsoids with half-axes 
$a=b=\sigma_d/(2\sqrt{q})$ and $c=q \sigma_d/2$. 

Our results confirm that polydispersity has the tendency to stabilize the
fluid-fluid coexistence region w.r.t.~crystallization. We have
discussed this trend, in agreement with earlier
studies,\cite{bib::meijer1994,bib::sear1997,bib::fasolo2005}
in Section~\ref{sec::sizepoly} for size polydisperse spherical polymer coils.
In Section~\ref{sec::shapepoly} we have shown that similar behavior can be
found for the case of shape polydisperse case. Upon increasing the degree of
polydispersity the fluid-fluid binodal ``crosses'' the fluid-solid coexistence
line and one obtains a stable critical point with associated liquid and gas
phases and a triple point. This effect seems to be robust against various ways
to incorporate polydispersity, which can be rationalized by the fact that
particles of a polydisperse depletion agent can fill the available free volume
in a system better than a monodisperse species.

Our main result is the study of mixtures of colloids and two species of
polydisperse (spherical) polymer coils. The polymer were described by a
bimodal distribution consisting of two polydisperse distributions. 
For the averaged size ratios of the polymer components chosen here, i.e. 
$\bar q =0.25$ and $\bar Q=2.0$, we find in distinct parts of the phase
diagram fluid-fluid phase separations in the two limiting cases $x=1$, when
only the smaller polymer species is present, and $x=0$, when only the larger
polymer species is present. For $x=0$ the corresponding fluid-fluid phase
separation is stable independent of the degree of polydispersity. In the
second limiting case, $x=1$, the fluid-fluid phase separation has to be
stabilized by a sufficient degree of polydispersity. We choose 
$z_{\bar q}=z_{\bar Q}=5$ and find for a range of the mixing parameter $x$ two 
stable liquid phases: beside the gas phase, which is diluted in the
density of colloids, we find a low density and a high density liquid
phase. The fluid-fluid coexistence region is w-shaped in this case and the
phase diagram exhibits two stable critical points and two triple points. 

For systems for which the inter-particle interactions contains two length
scales, such as short-range repulsion and long-range attraction,
Hemmer and Stell\cite{bib::hemmer1970,bib::stell1972} found that two stable
critical points may occur. More directly related to our situation is a system
of star polymers\cite{bib::loverso2003} in which similar features in the phase
diagrams are reported, as the effective interaction potential is comparable to
those of the model of Stell and Hemmer. For other model fluids similar effects
were observed.\cite{bib::franzese2001,bib::buldyrev2002,bib::malescio2002} We
point out that our findings are not restricted to a polydisperse depletion
agent. We find similar results for a ternary mixture in which we consider one
colloid species and two distinct monodisperse species of depletion agent.  

While often size or shape polydispersity is considered an undesirable
side-effect, originating from imperfections of particle synthesis, it is
possible to make use of this property by stabilizing fluid-fluid phase
separations. Moreover, the degree of polydispersity allows one to adjust the
phase diagram. From this point of view one might speculate about fine tuning
the phase behavior of colloidal mixtures by specifying the polydisperse
distribution of the depletion agent.

\begin{appendix}

\section{Explicit expressions for $\alpha_{\textrm{eff}}$}  \label{app::explicitExpressions}

Explicit expressions for the effective free-volume fraction in the case of
infinitely thin needles for the Schulz, Gaussian and
Hat-like distributions, Eqs.~(\ref{eqn::schulzDist})-(\ref{eqn::hatlikeDist}),
read
\begin{widetext}
\begin{eqnarray}
\alpha_{\textrm{eff},S}(\eta_c;\bar q,z) &=& {\frac {{2}^{z}{z}^{z}}{ \left( \bar q{\it \kappa}\,{\it R_c}+2\,z \right) 
^{z}{e^{{\it \bar \kappa}}}}},                                                                                                                                           \\
\alpha_{\textrm{eff},G}(\eta_c;\bar q,z) &=& {\exp{\left(- \bar \kappa-\bar q \kappa R_c/2 + 
{\frac{\kappa^2 R_c^2 }{16 z^2}}\right)}} \nonumber \\
&& \times \frac{1}{\erf(z \bar q) + 1}\left( 1 - {\it \erf} \left( {\frac {-{z}^{2}\bar q+{\it \kappa}\,{\it R_c}/4}{z}} \right) \right) ,      \\
\alpha_{\textrm{eff},G}'(\eta_c;\bar q,z) &=& {\exp{\left(- \bar \kappa-\bar q \kappa R_c/2 + 
{\frac{\kappa^2 R_c^2 }{16 z^2}}\right)}},                                                                                                                       \\
\alpha_{\textrm{eff},H}(\eta_c;\bar q,z) &=& \frac{2 z}{\kappa R_c} 
\exp{\left(-\bar \kappa -\bar q \kappa R_c/2\right)}
\sinh \left( {\frac {{\it \kappa}\,{\it R_c}}{2 z}} \right).
\end{eqnarray}
\end{widetext}
Since only $\kappa$ and $\bar \kappa$ enter these expressions for a
one-dimensional depletion agent, there is no difference between an approach
based on the PY or the BMCSL equation of state. In terms of the colloid packing
fraction $\eta_c$, average asymmetry ratio $\bar q$ and the polydispersity
parameter $z$ we obtain
\begin{widetext}
\begin{eqnarray}
\alpha_{\textrm{eff},S}(\eta_c;\bar q,z) &=& (2 z)^{z} \left( 1-{\it \eta_c} \right)  \left( 3\,{
\frac {\bar q{\it \eta_c}}{1-{\it \eta_c}}}+2\,z \right) ^{-z},                                          \\
\alpha_{\textrm{eff},G}(\eta_c;\bar q,z) &=& 
(1-\eta_c) \exp{\left(  -{\frac {3\bar q {\it \eta_c}}{2(1-{\it \eta_c})}}+ {\frac {9{{\it \eta_c}}^{2}}{ 16 z^{2}\left( 1-{\it \eta_c} \right) ^{2}}} \right)}                                                                                                                         \nonumber\\
& & \times \frac{1}{\erf(z \bar q) + 1} \left( 1 - { \it \erf} \left(  -z \bar q+{\frac {3 {\it \eta_c}}{4 z (1-{\it \eta_c})}} \right) \right),               \\
\alpha_{\textrm{eff},G}'(\eta_c;\bar q,z) &=& (1-\eta_c) \exp{\left(  -{\frac {3\bar q {\it \eta_c}}{2(1-{\it \eta_c})}}+ {\frac {9{{\it \eta_c}}^{2}}{ 16 z^{2}\left( 1-{\it \eta_c} \right) ^{2}}} \right)},                                                                   \\
\alpha_{\textrm{eff},H}(\eta_c;\bar q,z) &=& \frac{2 z (1-\eta_c)^2}{3 \eta_c} \exp{ \left( {{\frac {-3 \bar q{\it \eta_c}
}{2(1-{\it \eta_c})}}}\right)} \sinh \left( {\frac {3{\it \eta_c}}{2 z\, \left( 1-{
\it \eta_c} \right) }} \right).
\end{eqnarray}
\end{widetext}

%

In the case of infinitely thin disk- or hexagon-shaped platelets the expressions for the effective
free-volume fraction within the BMCSL approach are given by
\begin{widetext}
\begin{eqnarray}
\alpha_{\textrm{eff},S}(\eta_c;\bar q,z) &=& \left(\frac{z}{\bar q}\right)^z ( -2 \chi_2 )^{-z/2} 
    \exp{ \left( -\frac{\chi_1^2}{8 \chi_2} + \chi_0 \right) }
        D_{-z} \left( \frac{-\chi_1}{\sqrt{- 2 \chi_2}} \right),                 \\
\alpha_{\textrm{eff},G}(\eta_c;\bar q,z) &=& \exp{\left( {\frac {{z}^{2}{\bar q}^{2} \chi_2+ \chi_0
\,{z}^{2}- \chi_0\, \chi_2+{z}^{2}\bar q \chi_1+{ \chi_1}^{2}/4}{{z}^{2}- \chi_2}} \right) }                                        \nonumber\\
& & \times\left( {\it \erf} \left( {
\frac {{z}^{2}\bar q+ \chi_1/2}{\sqrt {{z}^{2}- \chi_2}}} \right) 
+1 \right) {\frac {z}{2 [\erf(z \bar q) + 1] \sqrt {{z}^{2}- \chi_2}}},                                                \\
\alpha_{\textrm{eff},G}'(\eta_c;\bar q,z) &=& \exp{ \left( {\frac {{z}^{2}{\bar q}^{2} \chi_2+ \chi_0
\,{z}^{2}- \chi_0\, \chi_2+{z}^{2}\bar q \chi_1+{ 
\chi_1}^{2}/4}{{z}^{2}- \chi_2}}\right)} {\frac {z}{\sqrt {{z}^{2}- \chi_2}}},                                                \\
\alpha_{\textrm{eff},H}(\eta_c;\bar q,z) &=& \frac{z}{4}\sqrt{\frac{\pi}{-\chi_2}}
{\exp{\left( \chi_0 - \frac{\chi_1^2}{4 \chi_2} \right)}}
\sum_{n=1}^{2} (-1)^n \erf \left( {\frac { 2 \chi_2( \bar q - (-1)^{n} )+\chi_1z}{2 z \sqrt {-\chi_2}}}  \right),
\end{eqnarray}
\end{widetext}
where $\chi_2 = -\beta \gamma a_d /q^2$, $\chi_1 = - \beta \kappa c_d / q$ and
$\chi_0 = - \beta \bar \kappa X_d$. The geometrical measures $a_d$, $c_d$ and
$X_d$ are given in Tab.~\ref{tab::geometricMeasures}. $D_{-z}$ with $z>0$ are
parabolic cylinder functions. 

\end{appendix}

%

\end{document}